\documentclass[letterpaper,twocolumn,english,aps,prl,superscriptaddress,showpacs]{revtex4}
\usepackage[T1]{fontenc}
\usepackage[latin9]{inputenc}
\usepackage{babel}
\usepackage{array}
\usepackage{booktabs}
\usepackage{amstext}
\usepackage{amssymb}
\usepackage{graphicx}
\usepackage[unicode=true,pdfusetitle,
 bookmarks=true,bookmarksnumbered=false,bookmarksopen=false,
 breaklinks=false,pdfborder={0 0 1},backref=false,colorlinks=false]
 {hyperref}

\makeatletter

\pdfpageheight\paperheight
\pdfpagewidth\paperwidth

\providecommand{\tabularnewline}{\\}

\@ifundefined{textcolor}{}
{%
 \definecolor{BLACK}{gray}{0}
 \definecolor{WHITE}{gray}{1}
 \definecolor{RED}{rgb}{1,0,0}
 \definecolor{GREEN}{rgb}{0,1,0}
 \definecolor{BLUE}{rgb}{0,0,1}
 \definecolor{CYAN}{cmyk}{1,0,0,0}
 \definecolor{MAGENTA}{cmyk}{0,1,0,0}
 \definecolor{YELLOW}{cmyk}{0,0,1,0}
}

\usepackage{placeins}

\usepackage{lineno}

\usepackage{multirow}

\usepackage[font=small,labelsep=period,singlelinecheck=false,justification=RaggedRight]{caption}
\makeatother

\begin{document}

\title{Continuous all-optical deceleration and single-photon cooling of
molecular beams}

\author{{A.M. Jayich}}

\affiliation{{UCLA Department of Physics and Astronomy, Los Angeles, CA 90095, USA}}

\author{{A.C. Vutha}}

\affiliation{{York University, Department of Physics and Astronomy, Toronto ON M3J 1P3, Canada}}

\author{{M.T. Hummon}}

\affiliation{{NIST and University of Colorado, Boulder, Boulder, CO 80309, USA}}

\author{{J.V. Porto}}

\affiliation{{National Institute of Standards and Technology, Gaithersburg, MD 20899, USA}}

\author{{W.C. Campbell}}

\affiliation{{UCLA Department of Physics and Astronomy, Los Angeles, CA 90095, USA}}
\begin{abstract}
Ultracold molecular gases are promising as an avenue to rich many-body
physics, quantum chemistry, quantum information, and precision measurements.
This richness, which flows from the complex internal structure of
molecules, makes the creation of ultracold molecular gases using traditional
methods (laser plus evaporative cooling) a challenge, in particular
due to the spontaneous decay of molecules into dark states. We propose
a way to circumvent this key bottleneck using an all-optical method
for decelerating molecules using stimulated absorption and emission
with a single ultrafast laser. We further describe single-photon cooling
of the decelerating molecules that exploits their high dark state
pumping rates, turning the principal obstacle to molecular laser cooling
into an advantage. Cooling and deceleration may be applied simultaneously
and continuously to load molecules into a trap. We discuss implementation
details including multi-level numerical simulations of strontium monohydride
(SrH). These techniques are applicable to a large number of molecular
species and atoms with the only requirement being an electric dipole
transition that can be accessed with an ultrafast laser.

\end{abstract}
\maketitle
Research with ultracold ($<1$ mK) atoms has greatly expanded our
knowledge of quantum many-body physics \citep{Bloch:2012fk}, precision
metrology \citep{Ockeloen:2013cr}, possible time- and space-variation
of fundamental constants \citep{Lea:2007oq}, and quantum information
science \citep{Garcia-Ripoll:2005nx}. Ultracold molecules are desirable
as a powerful extension of these efforts, but also as a promising
starting point for entirely new investigations. The simplest molecules,
diatomics, have more internal degrees of freedom than the two constituent
atoms. This makes the body-fixed electric dipole moment of polar molecules
accessible with laboratory electric fields, and could lead to discoveries
and insights exceeding the rich physics explored with atoms. These
oriented electric dipoles can be used to generate strong, long-range,
anisotropic dipole-dipole interactions \citep{YanNATURE13}, creating
new quantum simulators \citep{Micheli:2006fk}, and opening new avenues
in quantum computation \citep{DeMille:2002uq}, physical chemistry
\citep{Krems:2008gf,Bell:2009fk,Dulieu:2011uq}, and other fields
of physics \citep{Carr:2009fk}. In analogy with atoms, we anticipate
that a method to apply strong optical forces and to cool many types
of molecules will be a valuable resource in the emerging field of
ultracold molecules.

Historically, a necessary ingredient for optical control of the motion
of atoms and molecules has been an effective electronic cycling transition,
where spontaneous emission from the excited state populates only the
original ground state. Such transitions have been employed with great
success in Doppler cooling, Zeeman slowing, magneto-optical trapping,
and a host of other techniques that are the first steps in almost
every experiment utilizing ultracold atoms \citep{WinelandPRL78,Andreev:1981kx,Phillips1982,Raab1987}.
Spontaneous decay processes that lead to ``dark states'' that are
excluded from this cycle are present in most atoms, and are ubiquitous
in molecules due to their vibrational degree of freedom. One way to
circumvent this issue is to apply more lasers to reconnect these dark
states to the cycling transition \citep{Rosa:2004fk}. This approach
was recently demonstrated with carefully-chosen molecular species
\citep{Shuman:2010fk,Zhelyazkova:2013fk,Hummon:2013uq}, and shows
particular promise as a method to reach the ultracold regime for those
species. An unfortunate consequence of the dark state repumping schemes
utilized with these molecules is that they reduce the total scattering
rate with each additional repump. This can suppress the maximum achievable
optical force by one to two orders of magnitude. For the vast majority
of molecules, this force reduction is a serious constraint for beam
deceleration using Doppler forces, and would result in cumbersome
stopping lengths that will limit the cold molecule flux due to transverse
spreading, and the need to repump in a Doppler-insensitive manner
over a long interaction length.

We present here an alternative solution to the dark state problem
that utilizes \emph{stimulated} emission to prevent spontaneous emission
into dark states. This process exploits chirped picosecond pulses
to deterministically drive the excited molecule back to the original
ground state, eliminating the complications due to the multi-level
structure of molecules by isolating a 2-level system. This results
in a conservative force \citep{Kazantsev:1974ys,Voitsekhovich:1994kx,Nolle:1996fk}
that is considerably stronger than the Doppler cooling force. Ultrafast
stimulated slowing should be well-suited for the deceleration of molecules
from demonstrated molecular beam velocities to a full stop in the
lab. We then discuss how this deceleration technique (as well as most
others that have been demonstrated) can be augmented by a single-photon
velocity-cooling process that cools the decelerating molecules into
a continuous source of cold, trappable molecules. A few-photon trap
loading step such as demonstrated by Lu \textit{et al.} \citep{Lu:2013hc}
may also be added subsequently to compress the position distribution.
Molecules that have been slowed or trapped can then potentially be
cooled to the ultracold regime through sympathetic \citep{Myatt:1997fk},
evaporative \citep{Ketterle:1996pi}, or Doppler cooling. This approach
promises to extend the reach of laser cooling to molecules that are
considered difficult to directly laser cool -- a class that includes
many molecules that are interesting for applications such as precision
measurements \citep{Wilkening:1984uq,Hinds:1997lq,Vutha:2010vn} and
cold chemistry \citep{Krems:2008gf,Bell:2009fk,Dulieu:2011uq}.

\section{Ultrafast Laser Deceleration Force}

The ultrafast stimulated slowing we present here derives its mechanical
effect from momentum transfer between photons and molecules. The force
is generated by the fast repetition of a cycle where the molecules
are first illuminated by a ``pump pulse'' that is followed immediately
by a ``dump pulse''. The pump pulse counter-propagates with respect
to the molecular beam, and the absorption of a photon from this pulse
reduces each molecule's momentum by $\hbar k$, where $\hbar$ is
the reduced Planck constant and $k$ is the wave-vector. The molecules
are then deterministically driven back to their ground state by a
co-propagating dump pulse, which stimulates emission from each molecule,
removing another $\hbar k$ of forward momentum. This cycle can then
be rapidly repeated many times to produce a strong time-averaged continuous
deceleration force. Time-ordering of the pulses determines the sign
of the force, and occasional delays can be used to re-initialize the
populations, which happens quite naturally when generating the ultrafast
stimulated slowing from a standard $\sim$80 MHz repetition rate ultrafast
laser. In the limit that each laser pulse achieves full population
transfer, the optical comb tooth structure of the spectrum that arises
from inter-pulse phase coherence becomes irrelevant, and mode-locking
serves only as a convenient method for producing picosecond pulses.
While we propose a variation of this scheme utilizing chirped picosecond
pulses, this force has been used with un-chirped pulses to deflect
molecular \citep{Voitsekhovich:1994kx} and atomic \citep{Nolle:1996fk,Goepfert:1997vn}
beams in the transverse direction. In this mode of operation, it bears
some similarity to the bichromatic force \citep{SodingPRL97,YatsenkoPRA04,EylerArXiv}.
This un-chirped realization may be regarded as being (in some respects)
the polychromatic limit of the bichromatic force \citep{Galica:2013fk}.

The pulse duration of a few picoseconds is chosen to provide many
of the desirable features of this method. Picosecond pulses are much
shorter than the spontaneous emission lifetime, therefore the delay
between absorption and stimulated emission can be made short enough
that intervening spontaneous emission is negligible. This fast timescale
also results in the ability to repeat the cycle much faster than the
spontaneous scattering rate (which limits Doppler cooling), and we
estimate below that stopping distances on the order of 1 cm will be
attainable. The bandwidth of picosecond pulses is also much larger
than Doppler shifts for the entire velocity range from beam velocities
to a full stop in the lab, which results in a capture range that exceeds
room temperature. Further, as we discuss below, ultrafast pulses can
be chirped over large frequency ranges with passive optics to enhance
the state transfer fidelity via adiabatic rapid passage. Since the
bandwidths of picosecond pulses are also much smaller than femto-
or atto-second pulses, this allows one to limit undesired single and
multi-photon transitions (an important concern with molecules) through
frequency selection. Finally, a technological advantage of using ultrafast
lasers is that the pulses can be frequency-doubled with high efficiency,
thereby accessing a large variety of atoms and molecules that includes
those with transitions deep in the ultraviolet.

\section{Enhanced Fidelity Through Chirped Pulses\label{sec:Fidelity}}

Ultrafast stimulated slowing requires that pulses drive the absorption
and stimulated emission events. To repeat the cycle many times, each
laser pulse must perform an operation equivalent to a $\pi$ rotation
on the Bloch sphere (a ``$\pi$-pulse'') for each molecule in the
beam. It is crucial to realize high fidelity population transfers
for a large fraction of the molecular beam to avoid unwanted spontaneous
emission and to generate a large deceleration force. The fidelity
of population transfer by a transform-limited picosecond pulse will
inevitably be less than 1 for a large fraction of molecules in a molecular
beam for a host of reasons. The main source of decreased fidelity
is the laser's Gaussian spatial profile, causing a Rabi frequency
variation across the laser beam's transverse and longitudinal intensity
distribution. Shot-to-shot pulse energy variation is also a concern,
as is beam-pointing stability and a potential imbalance between the
co- and counter-propagating laser beams. Circumventing such sources
of population transfer infidelity is necessary if large numbers of
molecules are to be decelerated with pulses from a mode-locked laser,
as pointed out by Galica \textit{et al.} \citep{Galica:2013fk}.

Population transfer infidelity can be mitigated by the use of adiabatic
rapid passage (ARP) \citep{LoyPRL74,2001ARPC...52..763V}, where sweeping
the laser frequency through resonance can result in nearly 100\% transfer
fidelity. This technique has been examined for atomic deceleration
by Metcalf and co-workers \citep{Lu2005,LuPRA07,StackPRA11}, who
have employed long (multi-nanosecond) pulses and fast phase modulators
to drive ARP \citep{MetcalfDAMOP}. We note here an advantage of using
ultrafast lasers, where it is straightforward to ``chirp'' a transform-limited
pulse from an ultrafast laser by applying group delay dispersion (gdd)
either with a pair of gratings \citep{Treacy:1969fk}, chirped mirrors,
optical fibers \citep{Melinger1992}, or a Gires-Tournois interferometer
\citep{Kuhl:1986ys}. A chirped pulse has a time-varying frequency
that can sweep over a vast frequency range to drive ARP for all velocity
classes simultaneously. This chirp is particularly important for molecules
because in a multilevel system the chirped pulses can become state-selective
despite their large bandwidth \citep{Melinger1992}, effectively reducing
the complex molecular structure to a set of 2-level systems. 

\begin{figure}[h]
\includegraphics[width=1\columnwidth]{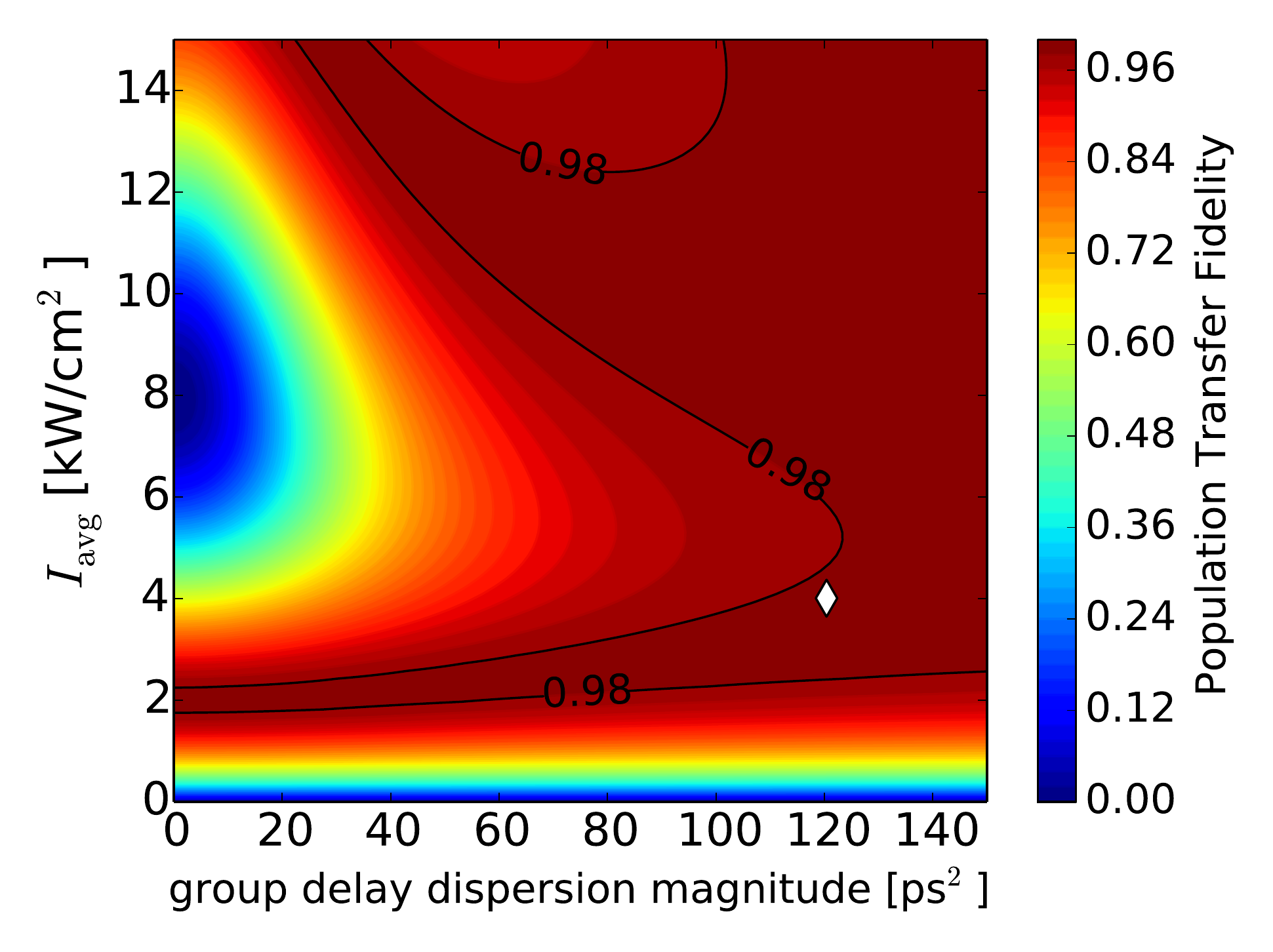}\caption{Single-pulse population transfer fidelity for SrH as a function of
the time-averaged laser intensity, $I_{\mathrm{avg}}$, and the group
delay dispersion (gdd) applied to a $\tau=7$ $\mathrm{ps}$ (temporal)
Gaussian transform-limited pulse. The calculated pulse is composed
of 99\% $\pi$-polarized light that connects the two sublevels of
an $X$-state $J=\frac{1}{2}$ manifold to two sublevels of an $A$-state
$J'=\frac{1}{2}$ manifold along a $Q$-branch. The hyperfine structure
is unresolved and therefore not included in the calculation. The white
diamond highlights the maximum intensity realized and the gdd used
in the Monte Carlo simulation shown in Figure \ref{fig:A-Monte-Carlo}.
The typical Rabi flopping as a function of intensity is realized for
$\mathrm{gdd}=0$.}

\label{fig:RAP fidelity}
\end{figure}

For ARP, it is important to maintain the adiabaticity criterion while
sweeping over a wide-enough frequency range to make the transfer robust
\citep{2001ARPC...52..763V}. Figure \ref{fig:RAP fidelity} shows
the results of a numerical calculation of the ARP population transfer
fidelity for SrH (see inset levels of Figure \ref{fig:2Pi - 2Sigma+}).
We plot the probability of populating the excited state (from an initial
ground state) as a function of the laser intensity and the gdd applied
to a transform-limited $\tau=7$ $\mathrm{ps}$ Gaussian pulse. The
operating point (highlighted with a white diamond in Fig. \ref{fig:RAP fidelity})
corresponds to 4 W of average power focused to a spot with an intensity
full width at half maximum of 0.3 mm. A grating pulse stretcher \citep{Treacy:1969fk}
spaced by 8.7 m with gratings of 2000 lines/mm can achieve -120 $\mathrm{ps}^{2}$
of gdd. This creates a high-fidelity laser-molecule interaction volume
where variations of the intensity as large as a factor of 2 will have
a negligible effect on the population transfer. The calculation shown
in Fig. \ref{fig:RAP fidelity} includes 1\% of the laser power having
undesired polarization to model the effect of inevitable imperfections
in the optics. These numerical calculations are used as inputs to
the Monte Carlo simulation shown in Fig. \ref{fig:A-Monte-Carlo}.


\section{Molecular Structure}

For the picosecond-scale pulses considered here, the pulse bandwidth
is much larger than hyperfine structure, and we consider only $J$
($N\text{, }v\text{, }F\text{, }$ and $J$ are good quantum numbers
in Hund's coupling case (b), as defined in \citep{Brown:2003oq}).
Since each ground quantum state that can be slowed requires a unique
excited quantum state for ultrafast stimulated slowing, the number
of ground states that can be simultaneously slowed on a given spectroscopic
transition is limited to either the multiplicity of the excited or
ground state (whichever is smaller). In order to decelerate all of
the population in the ground state, it would therefore seem desirable
to work on a transition such that these are equal ($J=J'$) by slowing
on a $Q$-branch transition with $\pi$-polarized pulses (a prime
denotes the excited state). However, for molecules with integer values
of $J$ (such as spin-singlets and triplets), $Q$-branch transitions
will have dark states even for pure $\pi$-polarization \citep{Berkeland:2002kx},
and an $R$-branch transition is desirable instead ($J=J'-1$). We
note here that driving a $P$-branch transition ($J=J'+1$, as is
done for molecular Doppler cooling) could potentially eliminate the
need to repump rotational branching, but would come at the cost of
dark ground-state sublevels on the pulsed transition, leading to ground-state
molecules that do not get decelerated and velocity, position, and
optical phase sensitivity that will likely degrade the population
transfer fidelity. Slowing on a $Q$- or $R$-branch transition may
require another laser to rotationally repump molecules that happen
to suffer the occasional spontaneous emission event, but this laser
may be applied transverse to the molecular beam to avoid Doppler shifts
and can be derived from either a pulsed or CW laser. Every increase
in population transfer fidelity reduces the reliance on repump lasers.

\begin{figure}[h]
\includegraphics[width=1\columnwidth]{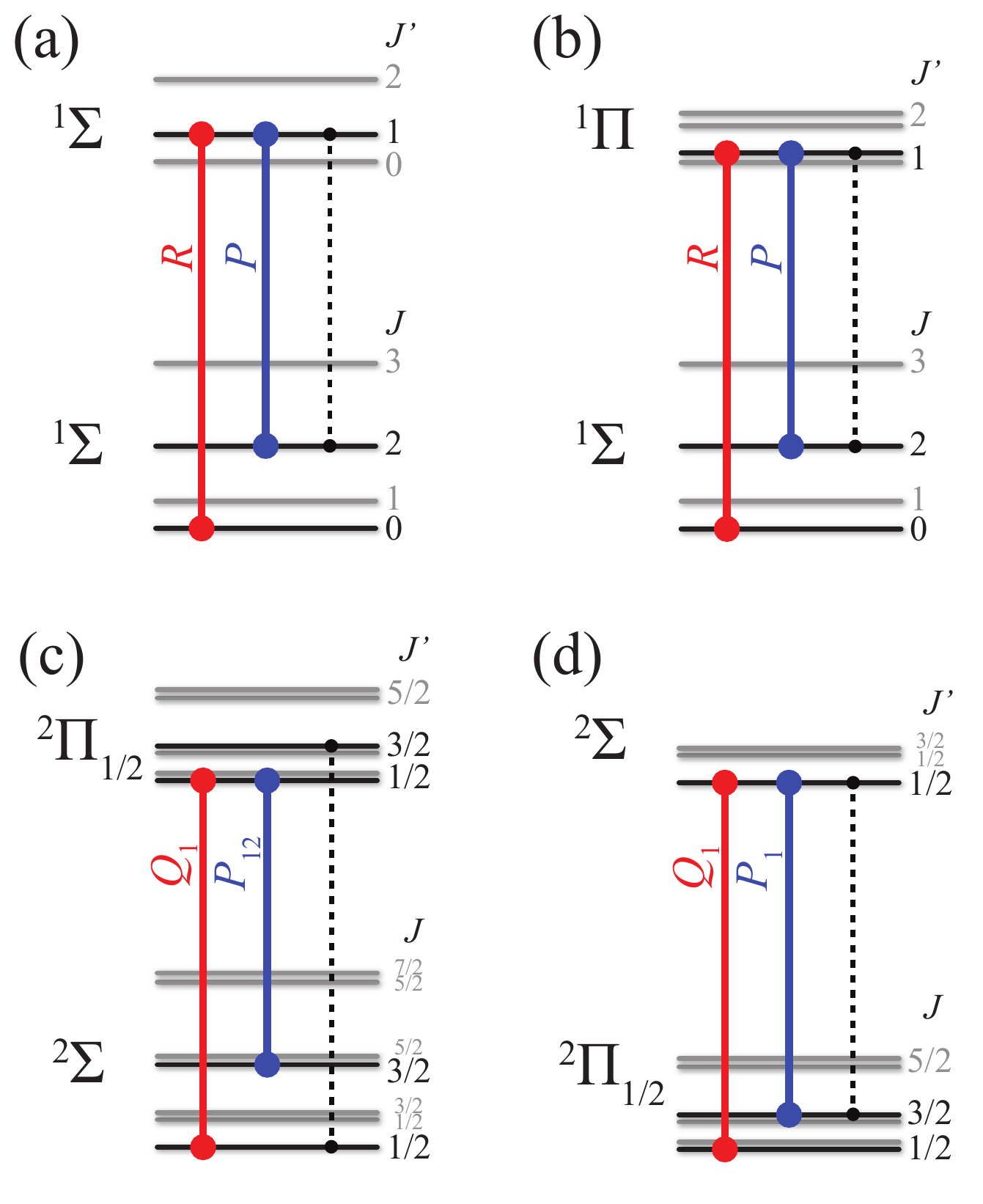}\caption{Example schemes for applying ultrafast stimulated slowing to ground-state
diatomics. The red lines show the deceleration transition, the blue
lines a rotational repump and the broken black lines indicate the
nearest transition that will out-couple molecules from the deceleration
cycle if they are significantly within the bandwidth of the pulsed
laser. Notable spectroscopically-characterized species with these
structures include (a) SrO, ThO (b) AlCl, AlF (c) SrF, YO, CaF, SrH
(d) CH, OH, SiF}
\label{fig:4 transition cooling schemes}
\end{figure}

Figure \ref{fig:4 transition cooling schemes} shows examples of how
the deceleration could be applied to four elementary types of molecular
transitions (two singlet and two triplet), though it is generally
applicable to any molecule with a strong electronic transition that
is accessible to an ultrafast laser. In all of these cases, the pulse
bandwidth is limited to be no larger than a few times the rotational
constant, which leads to the few picoseconds to tens of picoseconds
regime for many diatomic molecules. Even if the unwanted transition
is within the laser bandwidth, the ARP process can be made state-selective
\citep{Melinger1992} and the unwanted transition can be avoided at
the cost of requiring more group delay dispersion to lengthen the
pulse chirp in time. Unwanted transitions may be avoided if they are
separated from the desired line by at least half of the pulse bandwidth.
Numerical calculations confirm that the chirp does not need to be
increased substantially from what would be required for robust 2-state
ARP. Slight polarization imperfections can eventually lead to dark
states for molecules decelerated on $Q$-branch transitions, but we
calculate that the polarization purity can easily be maintained at
a level where shelving into dark states is not an issue. For this
branch the pulse chirp sign can be identical for the pump and dump
pulses. 

We can make a simple estimate of the distance required to stop the
molecular beam from its initial velocity, $V_{0}$. For a molecule
with mass $m$ and laser repetition rate $f_{\text{rep}}$, we find
the stopping distance for perfect fidelity population transfer is
given by

\begin{equation}
l_{\text{s}}=\frac{mV_{0}^{2}}{4\hbar kf_{\text{rep}}}.\label{eq:analytic stopping distance}
\end{equation}
The stopping distances for several species are given in Table \ref{Stopping distances}
for $V_{0}=200$ meters/second (m/s) and $V_{0}=50$ m/s with $f_{\text{rep}}=80$
MHz. 

\begin{table*}[h]
\begin{tabular}{c>{\centering}m{4cm}>{\centering}m{4cm}ccc}
\toprule 
Species & Stopping distance (cm) \linebreak ($V_{0}=200$ m/s) & Stopping distance (cm) \linebreak ($V_{0}=50$ m/s) & $\lambda$ (nm) & mass (amu) & Transition\tabularnewline
\midrule
\midrule 
AlF & 0.3 & 0.02 & 2278 & 46 & $^{1}\Pi\leftrightarrow^{1}\Sigma$\tabularnewline
\midrule 
CaO & 1.5 & 0.10 & 866 & 56 & $^{1}\Sigma\leftrightarrow^{1}\Sigma$\tabularnewline
\midrule 
CH & 0.2 & 0.01 & 431 & 13 & $^{2}\Delta\leftrightarrow^{2}\Pi$\tabularnewline
\midrule 
Rb & 2.1 & 0.13 & 780 & 85 & $^{2}P_{3/2}\leftrightarrow{}^{2}S_{1/2}$\tabularnewline
\midrule 
SiF & 0.7 & 0.04 & 439 & 47 & $^{2}\Sigma\leftrightarrow^{2}\Pi_{1/2}$\tabularnewline
\midrule 
SrH & 2.1 & 0.13 & 751 & 89 & $^{2}\Pi_{1/2}\leftrightarrow^{2}\Sigma$\tabularnewline
\midrule 
TlF & 2.0 & 0.12 & 284 & 223 & $^{3}\Pi_{0}\leftrightarrow^{1}\Sigma$\tabularnewline
\bottomrule
\end{tabular}\caption{Predicted stopping distances for a molecular beam with initial velocities
of 200 m/s and 50 m/s with a laser imparting $2\hbar k$ per laser
pulse with $f_{\mathrm{rep}}=80$ MHz. For some of these species the
deceleration and repumping level structure is outlined in Figure \ref{fig:4 transition cooling schemes}.
It is important to note that for diatomic molecules with rotational
constants significantly smaller than the laser pulse bandwidth a more
complicated scheme might be required to avoid rotational branching.}
\label{Stopping distances}
\end{table*}


\section{Phase Space Compression}

In order to cool the slowed molecules, entropy must be removed. Ultrafast
stimulated slowing displaces the momentum distribution, but does not
alter its width, and is therefore a deceleration method, not a cooling
method. In order to extract entropy from the molecules, we propose
to use a continuous-wave (CW) laser to cool the slowed molecules via
a single spontaneous emission event. Such single-photon cooling was
first demonstrated in 1991 by Cornell, Monroe, and Wieman \citep{Cornell:1991oz},
and has been studied extensively in various forms \citep{Raymond-Ooi:2003rw,Narevicius:2009dp,Thorn:2009la,Aghajani-Talesh:2009ec,Riedel:2011xr,Falkenau:2011eu,Lu:2013hc}.
The scheme we describe here can be thought of as the momentum-space
equivalent of these position-space ideas, and position-sensitive versions
(such as that recently demonstrated for CaF molecules \citep{Lu:2013hc})
can be applied with this scheme to complete the phase space compression
in both directions.

Within the framework of ultrafast stimulated slowing, a delay can
periodically be inserted after a burst of deceleration cycles during
which a collinear CW laser illuminates the molecules. This narrow-band
CW laser is tuned to optically pump ground state molecules into a
long-lived dark state (such as a different vibrational level) via
a single spontaneously-emitted photon, where they remain and are no
longer addressed by the deceleration laser. The high likelihood of
optical pumping of this sort is precisely the issue that makes molecules
difficult to laser-cool in the first place, but it can be exploited
to our advantage to efficiently cool each molecule in a single spontaneous
emission event. Since this CW optical pumping step is sensitive to
the velocity of the molecules via their Doppler shift, molecules only
get pumped into dark states when they reach the desired target velocity.
The spontaneous emission of a photon during this optical pumping process
heralds the arrival of the molecule into the desired target velocity,
thereby extracting entropy from the molecule.

Given a long interaction time with the CW beam, the final velocity
width of the optically pumped molecules can in principle be reduced
to the recoil limit if the CW laser drives a narrow transition. However,
in practice the deceleration rate will set the final velocity width
well before the recoil limit is reached. The reason for this is that
the molecules must spend enough time in the vicinity of the target
velocity to absorb a photon from the CW beam, so larger deceleration
rates will require more CW laser power, and power broadening will
dictate the addressable velocity width. If the time-averaged force
from the ultrafast stimulated slowing (including delays for optical
pumping) is given by $F=\hbar kf_{\mathrm{rep}}$, a CW velocity-cooling
laser of similar wavelength will produce a final velocity distribution
with an effective temperature
\begin{equation}
k_{\mathrm{B}}T_{\mathrm{rep}}\geq2\pi\hbar f_{\mathrm{rep}}\label{eq:cooling limit}
\end{equation}
where $k_{\mathrm{B}}$ is the Boltzmann constant. This relationship
demonstrates the connection between the deceleration force and the
final effective temperature in this scheme.

After their longitudinal momentum space density has been compressed,
molecules can be subjected to a similar single-photon cooling step
in position-space through position-sensitive optical pumping into
a trap \citep{Cornell:1991oz,StuhlerPRA01,Riedel:2011xr,Lu:2013hc}.
In principle, each molecule needs to spontaneously emit only two photons
during the entire process from beam to trap, demonstrating the large
capacity each spontaneously-emitted photon has for carrying away entropy.

\section{Application To Strontium Monohydride}

\begin{figure}[h]
\includegraphics[bb=0bp 0bp 491bp 470bp,width=1\columnwidth]{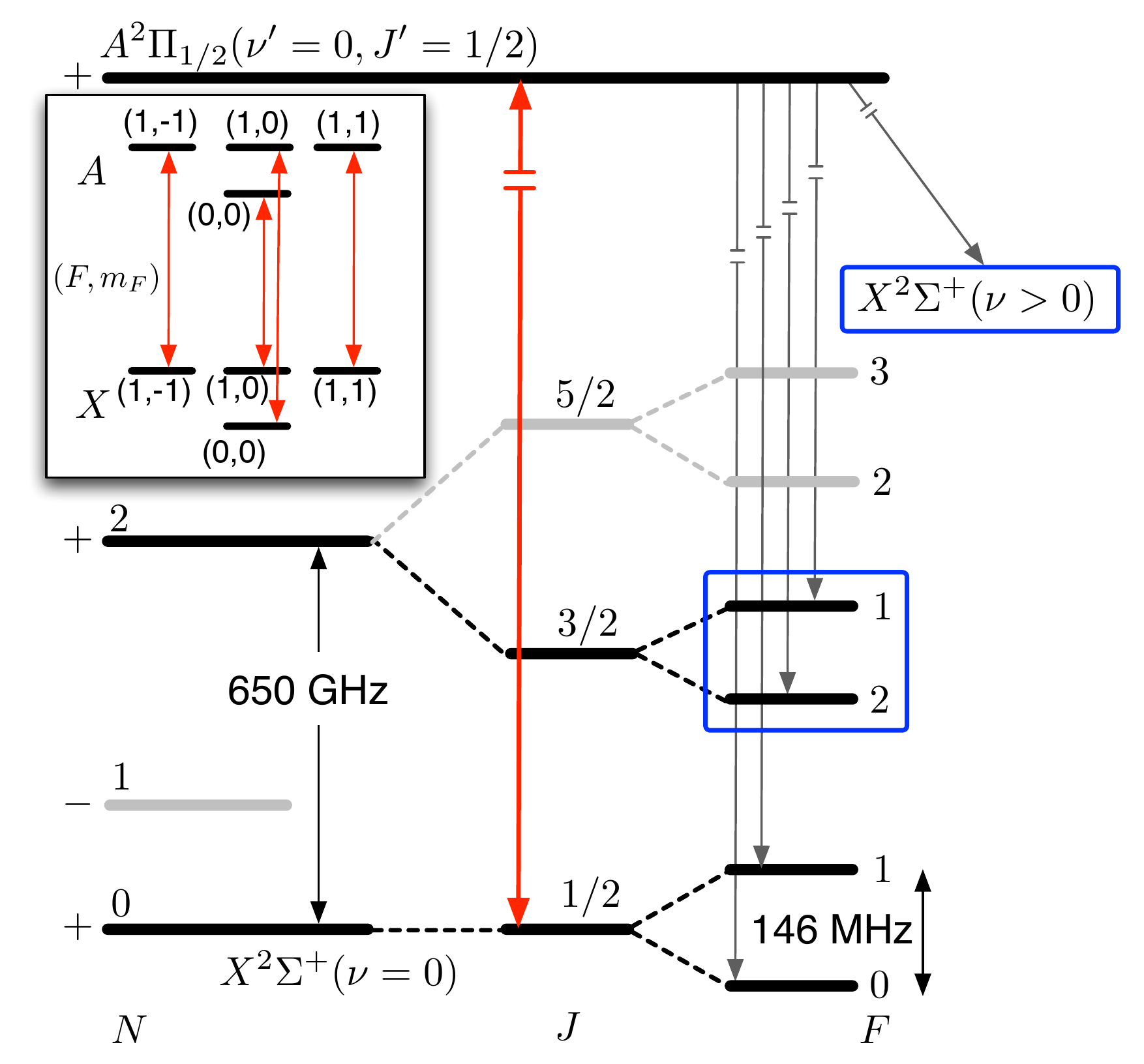}\caption{The pertinent level structure for $^{88}$SrH. A full slowing cycle
can be completed along the $Q_{1}$ branch, pictured as an inset,
with Zeeman sublevels, $m_{F}$, shown driven by $\pi$-polarized
light. Occasional decays from the excited state will populate the
desired slowing state, along with dark states highlighted in blue,
including higher vibrational states, which can be repumped with CW
or pulsed lasers. }
\label{fig:2Pi - 2Sigma+}
\end{figure}

We have used numerical simulations to investigate the performance
of the all-optical deceleration and cooling using $^{88}\mathrm{SrH}$
as a specific example. This molecule is considered for three primary
reasons: the $A^{2}\Pi_{1/2}\leftrightarrow X^{2}\Sigma^{+}$ electronic
transition is strong ($\tau=34$ ns), addressable by a Ti:Sapphire
laser ($\lambda=751$ nm), and has a low branching ratio to vibrational
dark states (1/67 \citep{Rosa:2004fk}) which makes it relatively
forgiving of population transfer infidelities. The relevant level
structure of $^{88}\mathrm{SrH}$ ($A^{2}\Pi_{1/2}\leftrightarrow X^{2}\Sigma^{+}$)
is depicted in Figure \ref{fig:2Pi - 2Sigma+}. We propose to execute
the deceleration cycle with $\pi$-polarized light from the $X$-state
$N=0$, $J=\frac{1}{2}$ manifold to the $A$-state $(N'=1)$, $J'=\frac{1}{2}$
manifold on the $Q_{1}$ branch. These two manifolds have the same
number of sublevels ($J=J'=\frac{1}{2}$), as shown in the inset of
Fig. \ref{fig:2Pi - 2Sigma+}, and $\pi$-polarized light will therefore
slow all molecules in the rotational ground state. 


\begin{figure}[h]
\includegraphics[clip,width=1\columnwidth]{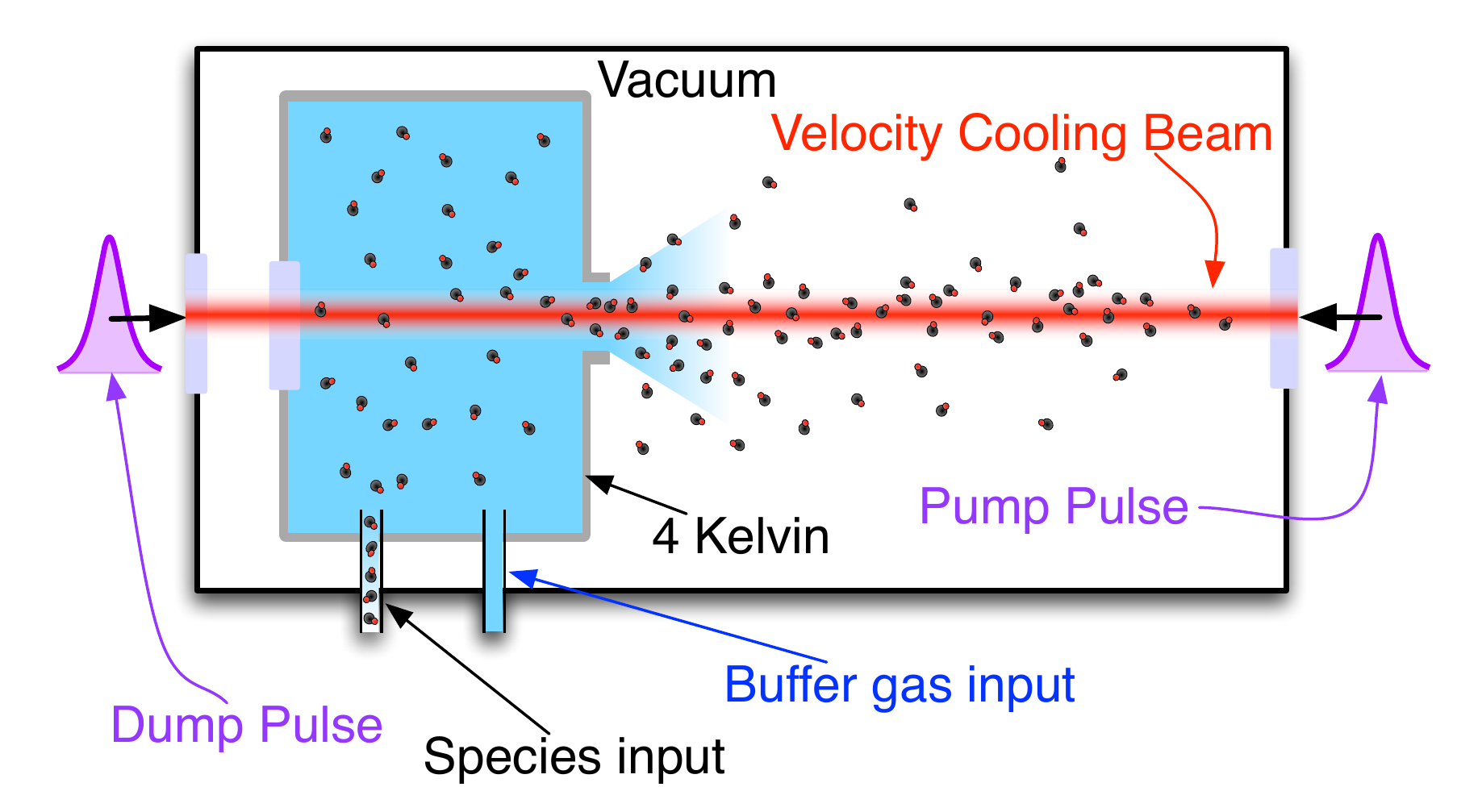}\caption{Molecules emitted from a cryogenic buffer-gas beam (CBGB) source are
in their ro-vibrational ground state. After exiting the nozzle they
are slowed by picosecond pulses (shown in purple). A velocity-selective
optical-pumping laser (red) drives single-photon cooling and compresses
the velocity-space density of the beam. }
\label{fig:Beam cartoon}
\end{figure}


To produce ro-vibrationally cold molecules we envision using a cryogenic
buffer-gas beam (CBGB) source, as depicted in Fig. \ref{fig:Beam cartoon},
such as those described in \citep{Hutzler2012}. In these CBGB sources,
molecules are cooled by collisions with a cold buffer gas (\emph{viz.}
helium or neon), and intense beams with forward velocities as low
as 35 m/s \citep{Patterson:2007bv} have been produced. 


\begin{figure*}[h]
\includegraphics[width=2\columnwidth]{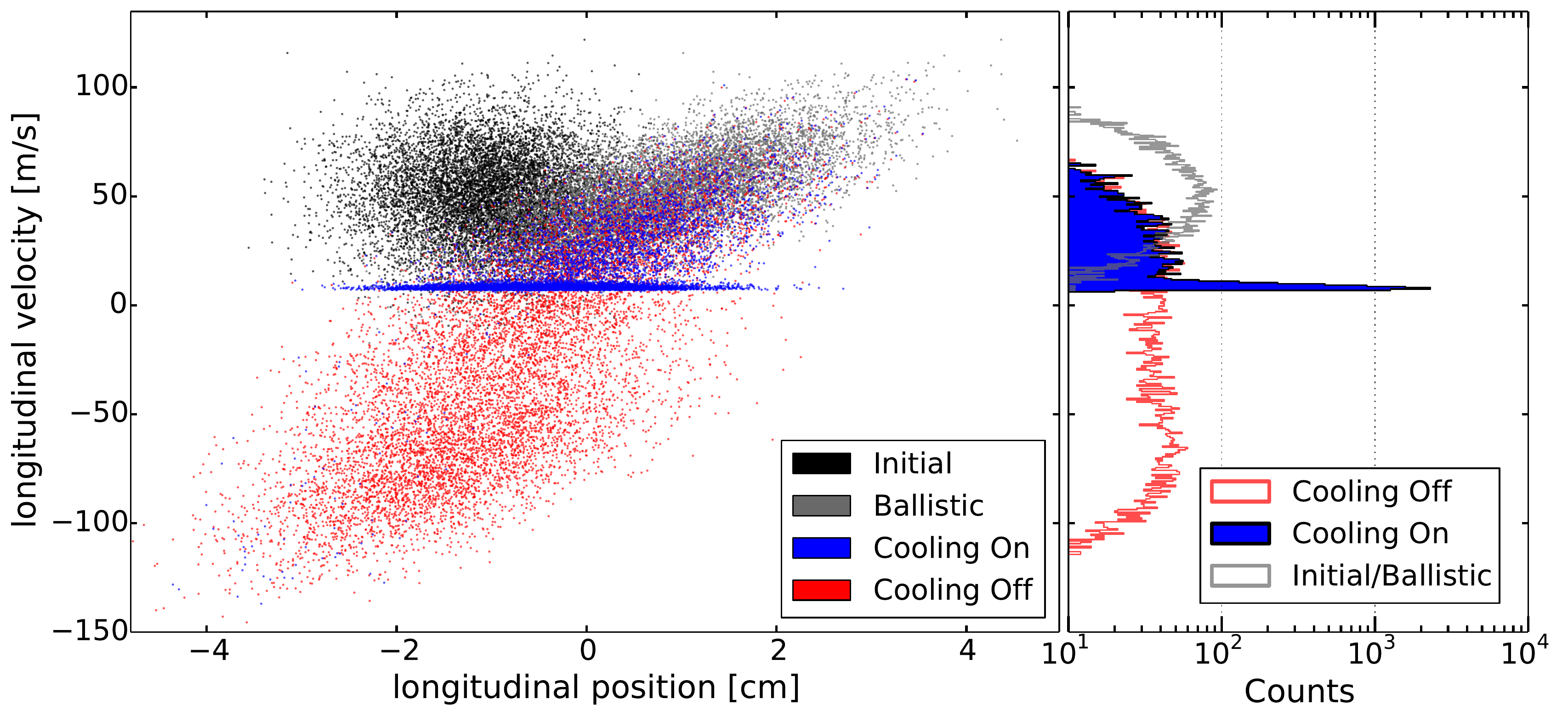}\caption{A Monte Carlo simulation of the deceleration and cooling for SrH.
The longitudinal phase space distribution is shown for the initial
distribution (black), ballistic trajectories (grey), the trajectories
when the ultrafast laser deceleration force is applied (red), and
the trajectories when both the deceleration force and single-photon-cooling
lasers are applied (blue). A histogram of longitudinal velocities
is shown on the right on a semilog scale. \label{fig:A-Monte-Carlo}}
\end{figure*}

Figure \ref{fig:A-Monte-Carlo} shows the computer simulation of the
longitudinal phase-space evolution obtained by tracking full 3-dimensional
molecule trajectories (for a detailed treatment of the simulation
parameters, see the appendix). The black distribution represents the
initial beam as it exits the CBGB source, which evolves into the grey
distribution in the absence of deceleration and cooling. The blue
points show the effect of deceleration by ultrafast pulses and cooling
by the CW velocity-cooling laser. The deceleration translates the
distribution toward negative velocity. We simulate deceleration cycle
bursts that are each 28 pulses long, interleaved with delay periods
of repumping and single-photon cooling. The duration of these delays
results in an effective time-averaged repetition rate of $f_{\mathrm{rep}}=47$
MHz. The single-photon cooling laser was tuned to be resonant with
molecules traveling with a longitudinal velocity near 7 m/s. Molecules
gather near this target velocity due to the optical pumping process,
and the histogram shown on the right side of Fig. \ref{fig:A-Monte-Carlo}
shows this density increase on a semilog scale. We find that molecules
with large transverse velocity wander out of the deceleration laser
beam and therefore never make it to the target velocity. With the
addition of a molecular guide (magnetic or electric), it may be possible
to enhance the range of transverse velocities and positions that are
captured.

For those molecules that were successfully optically-pumped out of
the deceleration cycle by the CW velocity-cooling laser (which constitute
more than half of those simulated), their momentum was reduced by
an average of $36\hbar k$ per spontaneously-emitted photon. This
value is $1\hbar k$ for traditional laser cooling. This demonstrates
the power of ultrafast stimulated slowing to open the door to species
with branching ratios tens of times worse than the special Doppler-coolable
species. Furthermore, the total distance required for this deceleration
is $\sim1$ cm, as shown by the \textit{x}-axis of Fig. \ref{fig:A-Monte-Carlo},
providing a simplification of the apparatus required to produce cold
molecules. We did not numerically simulate the addition of a position-selective
optical pumping laser or trap (which has been demonstrated recently
with molecules \citep{Lu:2013hc}), but we find that the velocity
compression alone increased the peak 6-dimensional phase-space density
by a factor of 16. The average stopping distance for the cooled molecules
is 0.4 cm. This agrees reasonably with the simple analytic expression
of Eq. \ref{eq:analytic stopping distance}, which gives 0.2 cm with
$f_{\mathrm{rep}}=47$ MHz (the effective repetition rate is reduced
by the cooling and repumping window).


\section{Summary}

We have outlined an all-optical scheme to generate trappable, cold
molecules that can be applied to a variety of molecules with the only
requirement being a transition that can be accessed with ultrafast
lasers. This opens up access to a very large number of diatomic molecules,
as well as atoms that are not traditionally amenable to laser cooling.
Simulations indicate that significant deceleration and cooling are
possible with existing technology. The technique presented here has
the desirable feature of being all-optical, circumventing the need
for complex \emph{in vacuo} components for cold molecule generation.
For this reason we anticipate that it will be sufficiently flexible
to be used in concert with other techniques that have been developed
over the past 15 years to produce ultracold molecules \citep{Bethlem1999,Rangwala:2003wd,Patterson:2007bv,Vanhaecke:2007eu,Zeppenfeld:2012fk}.
Finally, we note that another application of this optical force and
velocity-cooling process is that it can also be used to \emph{accelerate}
a sample of cold atoms or molecules to large velocities and produce
narrow distributions. Atomic beams with high spectral luminosity such
as this might be used to explore collisional physics, including narrow
shape resonances \citep{Boesten:1997fk} or to augment atom interferometry
\citep{Chiow:2011ly,Wacker:2010cr}.

\section*{ACKNOWLEDGMENTS}

The authors wish to thank Chris Monroe, Luis Orozco and Steve Rolston
for generous support of this project, as well as David DeMille, John
Doyle and Eric Hudson for helpful discussions. This work was supported
by a seed funding grant from the Joint Quantum Institute Physics Frontier
Center and the US Air Force Office of Scientific Research Young Investigator
Program under award number FA9550-13-1-0167.

\section*{APPENDIX}

The numerical simulation presented in Fig. \ref{fig:A-Monte-Carlo}
consists of two parts: the molecular beam source and the control lasers.
We combine estimates based on current technology for cold molecular
beam sources and picosecond lasers to estimate a number of decelerated
and cooled SrH molecules (which could be subsequently trapped). We
consider a molecular beam with center of mass velocity 50 m/s as in
\citep{Patterson:2007bv}, which is realizable for many species \citep{DoylePrivateCommunication}.
The simulation is performed assuming 4 W of time-averaged laser power
at a repetition rate of 80 MHz in 7 ps transform-limited (temporal)
Gaussian pulses that are chirped by -120 $\mathrm{ps}^{2}$ of gdd
before interacting with the molecules. The temporal Gaussian is chosen
for computational ease, and results with other similar pulse shapes
(such as hyperbolic secant) are similar. To determine the population
transfer fidelity we calculate the ARP dynamics for a given intensity
by time-evolving the Schr\"odinger equation with two ground states
and two excited states.

We assume that a single pulse is recycled, \emph{e.g.} by a mirror,
to generate the pump and dump pulses for a deceleration cycle. The
probability of population transfer by a chirped pulse is determined
by the gdd and the intensity, the latter of which varies as a function
of position. The state of the molecule is not evolved between pump
and dump pulses, as this delay can be made essentially as short as
the pulse widths, which are negligible compared to the spontaneous
emission lifetime. After the deceleration cycle the molecule evolves
ballistically for $1/f_{\mathrm{rep}}$. During this time if a molecule
is left in the excited state (due to bad fidelity) it may decay (every
spontaneous decay imparts a momentum kick in a random direction) back
to the ground state or to a dark state. The branching to the dark
state is given by the combination of the rotational branching to the
$N=2$, $J=3/2$ manifold (1/3) and to the $v=1$ states (1/67). Molecules
in the dark state are not decelerated. Because of intensity-variation-induced
population transfer infidelity, dark state repumping is necessary,
and after every 14 deceleration cycles (28 pulses) a repumping (and
cooling) step is inserted. We assume that 3/8 of the dark state molecules
are repumped to the excited state, from which they can decay again
to the ground or dark state with the branching ratios above. The decay
probability upon excitation by the repump is near unity because the
repumping and cooling window is 125 $\mu\mathrm{s}$. In order to
achieve a low final temperature, it is important to have a long cooling
window, as illustrated by Eq. \ref{eq:cooling limit}. The cooling
laser is detuned 9 MHz from the zero velocity class, which corresponds
to a forward velocity near 7 m/s. The cooling intensity (0.10 $\mathrm{mW}/\mathrm{mm}^{2}$)
is set so that a near $\pi$ rotation is performed on a $\gamma/2\pi=1$
MHz transition to an excited state that preferentially decays to a
state that is not addressed by the ultrafast laser. During the repumping
and cooling window the molecules evolve ballistically. The burst of
14 deceleration cycles (28 total pulses) followed by the 125 $\mu\mathrm{s}$
repumping and cooling window is repeated 1,200 times, giving a total
of 33,600 deceleration pulses.

Molecules with high transverse velocity will not be significantly
decelerated since they will quickly leave the high-intensity region
of space. Therefore we ignore any molecules that have a transverse
velocity greater than 0.75 m/s, giving rise to a simulated fraction
of $\sim8\cdot10^{-4}$. We use a diameter of 0.35 mm for the molecular
beam source. 

For the simulation shown in Fig. \ref{fig:A-Monte-Carlo}, the slowing
and cooling process captured more than 68\% of the molecules. For
these cooled molecules, an average of $36\hbar k$ of momentum is
transferred per (unwanted) spontaneous decay event. The phase space
of the cooled molecules was compressed by a factor of 16.1 (in other
simulations with hotter, faster samples the compression factor can
be as large as 35). Of the molecules that were cooled 94\% are trappable,
where a cooled molecule is considered trappable if it is moving $<10$
m/s and intersects a 1 cm diameter circle that is 2 cm from the source. 

\bibliographystyle{apsrev4-1}
\bibliography{Concept_Paper}

\end{document}